\def\cm{{\rm\thinspace cm}}

\def\erg{{\rm\thinspace erg}}

\def\K{{\rm\thinspace K}}
\def\keV{{\rm\thinspace keV}}
\def\km{{\rm\thinspace km}}
\def\kpc{{\rm\thinspace kpc}}

\def\Mpc{{\rm\thinspace Mpc}}
\def\Msun{\hbox{$\rm\thinspace M_{\odot}$}}

\def\s{{\rm\thinspace s}}

\def\pcmcu{\hbox{$\cm^{-3}\,$}}

\def\ergpcmsqps{\hbox{$\erg\cm^{-2}\s^{-1}\,$}}

\def\ergps{\hbox{$\erg\s^{-1}\,$}}

\def\kmps{\hbox{$\km\s^{-1}\,$}}

\def\pcmsq{\hbox{$\cm^{-2}\,$}}

\def\spose#1{\hbox to 0pt{#1\hss}}
\def\approxlt{\mathrel{\spose{\lower 3pt\hbox{$\sim$}}
        \raise 2.0pt\hbox{$<$}}}
\def\approxgt{\mathrel{\spose{\lower 3pt\hbox{$\sim$}}
        \raise 2.0pt\hbox{$>$}}}

\documentstyle[psfig]{mn}

\title[ADAF in M60]
{Advectively dominated flows in the cores of  giant elliptical galaxies:
 application to M60 (NGC 4649)}

\author[T.~Di~Matteo and A.~C.~Fabian]
{T.~Di~Matteo$^1$ and  A.~C.~Fabian$^1$\\ 
{\small $^1$Institute of Astronomy, Madingley Road,
Cambridge, CB3 OHA}\\ }

\begin{document}

\maketitle

\begin{abstract}
It has been suggested that the final stages of 
accretion in present--day giant elliptical galaxies occur in 
an advection-dominated
mode. The poor radiative efficiency of this accretion solution accounts 
for the fact that massive black holes, which are believed
to reside in the centres of these galaxies,
do not have the luminosities expected if accretion
from the hot interstellar medium occurs at the Bondi rate. 
We discuss the advection-dominated solution for the  nucleus of M60 (NGC 4649)
 and show that accretion at the Bondi rate is consistent
with the core flux from radio to X-ray. This solution  allows  
 for a black hole mass of $\sim 10^9\Msun$  as required by 
independent arguments. The successful application of this model to both 
 M60, and previously to the well known  nucleus of M87, suggests
that accretion of hot gas in an elliptical galaxy  creates the ideal 
circumstances for advection-dominated accretion flows to operate.

\end{abstract}


\section{Introduction}

Most nearby, large elliptical galaxies should  host a massive
black hole left over from an earlier quasar phase
(Fabian \& Canizares 1988; Fabian \& Rees 1995).
Giant elliptical galaxies also contain an extensive hot halo  
which provides a minimum fuelling level for accretion. This leads
to a problem. Fabian \& Canizares (1988) have determined a limit to the 
accretion rate for a sample of bright elliptical galaxies
 produced using the classical Bondi (1952) formula. 
They concluded that if the  radiative
 efficiency of accretion is $\sim 10$ per cent, then all such nuclei would 
appear much more luminous 
than  observed.
The limits on the luminosity of the galaxies mean that black holes
 masses are of at most a few times $10^7\Msun$ contradicting   
quasar counts and integrated luminosities  
 which imply black hole masses 
above $10^8$ -- $10^9\Msun$.
Low-luminosity radio activity is very common in ellipticals (Sadler, 
Jenkins \& Kotanij 1989; Wrobel 1991, Slee et al. 1994) indicating that 
indeed there is some kind of central engine in the cores of these galaxies.

The inconsistencies described above, can be reconciled by considering
a situation in which the intrinsic radiative efficiency of the
accretion is low. This has been noted by Fabian \& Rees (1995) in the
light of recent discussions of optically thin advection-dominated
accretion solutions (Narayan \& Yi 1995a,b; Abramowicz et al. 1995,
Narayan, Mahadevan \& Yi 1995 and references therein). In this mode of
accretion, the accretion rate, $\dot{M}$, is low, the ions remain at
the virial temperature and the radiative efficiency of the low density
accreting material can be very small. The energy released by viscous
friction is advected into the black hole.

Accretion of hot gas in an elliptical galaxy may create the ideal
circumstances for a hot ion torus to form (Begelman 1986) and thus an
advection-dominated accretion flow (ADAF) to operate \footnote{We note
that the gas in the central region of elliptical galaxies is very
likely to share angular momentum from stellar mass loss}. In
particular the successful application of low efficiency accretion via
an ADAF in the giant elliptical galaxy M87 (which contains a $3\times
10^9\Msun$ black hole -- Ford et al. 1995; Harms et al. 1995 ) by
Reynolds et al. (1997; Paper I) suggests that such mechanisms can be
of wide importance.  There, we found that the luminosity of M87 is
consistent with the observations only if Bondi accretion from the hot
gas involves an ADAF.  X-ray observations provide us with core gas
densities and therefore the means for estimating the Bondi accretion
rate.

Here we present another example of a giant elliptical galaxy, M60 (NGC
4649), in which we believe accretion could be taking place via an
ADAF. The giant elliptical galaxy M60 (NGC 4649) is an optically
luminous galaxy in the Virgo cluster. It possesses very weak jets and
small radio lobes indicating an active nucleus. The radio spectrum of
the nucleus is weakly inverted (Fabbiano et al. 1987) as expected from
an ADAF.

 A deprojection analysis of data from the
{\it ROSAT} High Resolution Imager (HRI) shows that the interstellar
medium (ISM) in M60 has a
central density $n\approx 0.1\pcmcu$ for a sound speed
$c_{\rm s}=300\kmps$ (C.~B.~Peres, private communication).  The
resulting Bondi accretion rate onto the central black hole is then
\begin{equation}
\dot{M}_{Bondi}=3\times10^3 \left(\frac{M}{\Msun}\right)^2\,\frac{n_{\infty}}
{c_{\infty}^3} \;\;\Msun {\rm yr^{-1}}.
\end{equation}
If we assume a black hole mass of $\sim10^9\Msun$ and a radiative efficiency
  of $\eta=0.1$, the Bondi rate for
a radiative efficiency of $\eta=0.1$ predicts a luminosity of the
order of $6\times10^{43}\ergps$. We show here that the 
luminosity of the core of M60 does not exceed $10^{41}\ergps$ 
and is probably less than $10^{40}\ergps$ over all energies.    

In this {\it Letter} we present a detailed examination of the
possibility that the massive black hole in M60 accretes via an ADAF.
In particular, we compute the spectrum expected and show that it is
consistent with  observations for a likely black hole mass 
($\approx 10^9$ \Msun) and  reasonable mass
accretion rates given by eqn.(1). 
In contrast to M87 (Paper I) the observations of M60 are not
highly contaminated by  jet emission and therefore  provide tighter
constraints on the model spectrum.

In Section 2 we compile data 
from the literature on
the full-band spectrum of the core of M60 and present some additional
data on the X-ray flux from the core. 
 Section 3 describes some details of our ADAF model spectrum
calculation. In  section 4, we compare the model spectrum with the data and
find that accretion rates comparable with the Bondi rate 
allow a black mass of $\sim 10^9\Msun$ at least.
\section{The spectrum of the core emission}

\subsection{The M60 data}

\begin{table*}
\caption{Summary of data for the core of M60.}
\begin{center}
\begin{tabular}{ccccc}\hline
Frequency & $\nu F_{\nu}$ & reference & notes \\
$\nu$ (Hz) & (10$^{-15}$\,erg\,s$^{-1}$\,cm$^{-2}$) & \vspace{0.3cm} \\ 
$1.4\times 10^9$ & 0.37 & Hummel et al. (1983) & Westerbork \\
$4.75\times 10^9$ & 1.14 & Fabbiano  et al. (1987), Wrobel (1991) & Effelsberg, VLA\\
$1.07\times 10^{10}$ & 2.78 & Fabbiano et al. (1987) & Effelsberg \\
$3.3\times 10^{10}$  & 10.5 & Fabbiano et al. (1987) & Effelsberg \\
$5.45\times 10^{14}$ & $\le$ 180 & Byun et al. (1996) & HST \\
$2.4\times 10^{17}$  & $\le$150 & this work & {\it ROSAT} HRI\\\hline

\end{tabular}
\end{center}
\end{table*}

 \begin{figure}
 \centerline{\psfig{figure=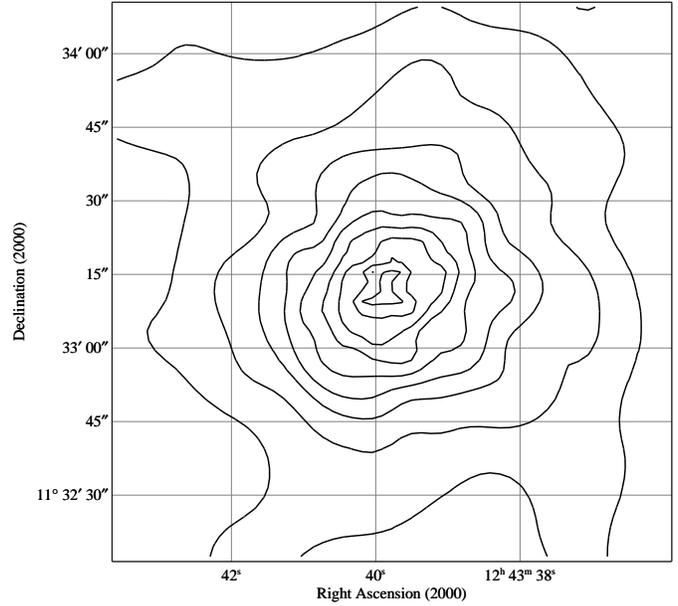,width=0.5\textwidth}}
 \caption{The core regions of M60 (NGC 4649) as imaged in a 20ks exposure with 
 the {\it ROSAT} HRI. The distribution of the X-ray emission in the central
 region is quite complex and asymmetric.  The diffuse emission is from
 the hot interstellar medium. The image has been smoothed adaptively prior to
 contouring; the minimum number of counts over which smoothing occurred is 25.
 Contour levels are equally spaced on a logarithmic scale, starting at 
$8\times10^{-2} {\rm count\; s^{-1} arcmin^{-2}}$ and increasing by a factor
1.6 between adjacent contours.}
 \end{figure}

\begin{figure}
 \centerline{\psfig{figure=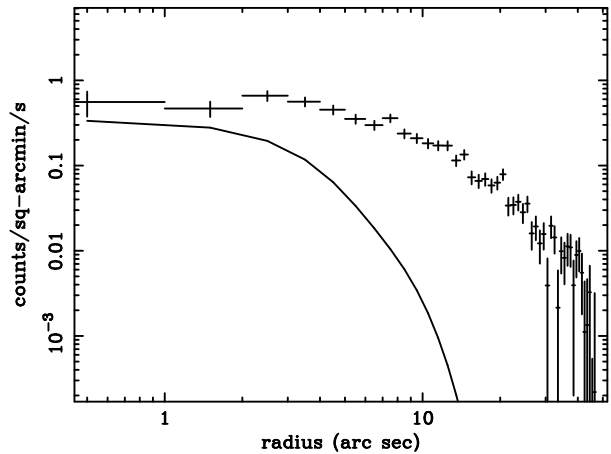,width=0.5\textwidth,angle=270}}
 \caption{The contribution of a point source to the emission of M60.
The solid line shows the PSF of the {\it ROSAT} HRI compared to the actual 
count rate per unit area.}
  \end{figure}

\noindent In order to examine the nature of the accretion flow in M60, we have
 compiled the best observational limits on the full band
spectrum of the core emission.  Our aim is to obtain good
observational limits on the core flux over a wide range of frequencies
rather than to compile a comprehensive list of all previous
observations. Some  contribution from the weak jets and in particular 
from the underlying galaxy
are unavoidable and so the derived spectrum should be considered an
upper limit to that of the accretion flow at the core of M60. The
data are summarized in Table~1.

Very Large Array (VLA) imaging by Stanger \& Warwick (1986) and Wrobel
 \& Heeshen (1991) shows an extended radio source dominated by a
 compact ($<$ 4 arc--sec) core.  The extended structure defines radio
 lobes powered by weak jets with linear dimensions $\sim 2 {\rm
 \kpc}$.  The slowly-rising spectral nature of the high-frequency
 emission is well defined by the data of Fabbiano et al. (1987)
 obtained from the 100 m radio telescope of the MPIfR at Bonn.

{\it Einstein} observations of M60 reveal that the overall
X-ray emission is extended with most of it arising from the hot
gas at ${\rm kT}=1{\rm \keV}$ (Stanger
 and Warwick 1986). 

We have examined the {\it ROSAT HRI} dataset in order to
constrain the nuclear X-ray flux of M60.  Here we  present the {\it ROSAT} HRI
 data resulting  from a 19\,923\,s exposure performed
on 1995 June 21.  Fig.~1 shows the {\it ROSAT} HRI image of M60.  
Consistently with the {\it Einstein} HRI observation, the
 {\it ROSAT} HRI image shows that M60 is extended X-ray source with most of the emission 
coming from hot gas in the interstellar medium. 
The low luminosity radio activity though, indicates that there is some form
of active central engine in the core of M60. 
In order to find an upper limit to the nuclear emission of M60, we have 
considered an extended component modelled
by a King profile plus a point source modelled by the PSF of the HRI 
(see Fig.2).
Our upper limit (at 90 per cent confidence) predicts a count rate from the 
point source component of $4.7\times 10^{-3}$\,ct\,s$^{-1}$ (Fig. 2). 
 Assuming the
spectrum to be a power-law with a canonical photon index $\Gamma=2.0$
modified by the effects of Galactic
absorption (with column density $N_{\rm H}=3.0\times 10^{20}\pcmsq$);
 this count rate implies a flux density at 1\,keV
of $F(1\,{\rm keV})=1.5\times10^{-13}\ergpcmsqps\keV^{-1}$.  This result is fairly 
insensitive to the choice of  power-law index.

\section{Advection-dominated tori}

It has been found (Rees 1982; Narayan \& Yi 1995ab; Abramovicz et al. 1995) 
that the advection dominated solution can occur only at accretion rates below
$\dot{M}_{\rm crit}\approx \alpha^2\dot{M}_{\rm Edd}$, where $\alpha$
is the Shakura-Sunyaev viscosity prescription and $\dot{M}_{\rm Edd}$
corresponds to the standard Eddington accretion rate.
At such low accretion rate the cooling timescale for the ions exceeds 
 the inflow timescale and most
 of the thermal energy is carried into the event horizon.
The plasma is much hotter than in the
classical, radiatively--efficient, thin disk solution and the flow
consists of a thick torus.
We assume that electrons and protons are coupled
only by two--body Coulomb interactions and that electrons radiate 
 by synchro-cyclotron,
bremsstrahlung and inverse Compton processes. Since the  Coulomb 
coupling is very 
weak, the ions remain at the virial temperature and very little energy
is transferred to the electrons. The disk, in this solution, has low
radiative efficiency.

For convenience, we rescale the radial co-ordinate and define $r$ by
\begin{equation}
r=\frac{R}{R_{\rm S}},
\end{equation}
where $R$ is the radial coordinate and $R_{\rm S}$ is the
Schwarzschild radius of the hole  
and the accretion rate,
\begin{equation}
\dot{m}=\frac{\dot{M}}{\dot{M}_{\rm Edd}}.
\end{equation}
The ADAF model, which is based on that of Narayan \& Yi (1995) is the
same as that used in Paper I for M87 (Reynolds et al. 1997).  By
assuming that the system is undergoing advection-dominated accretion,
we can predict the radio--to--X-ray spectrum of the accretion flow.
The amount of emission from the different processes and the shape of
the spectrum can be determined as a function of the model variables:
the viscosity parameter, $\alpha$, the ratio of magnetic to total
pressure, $\beta$, the mass of the central black hole, $M$, and the
accretion rate, $\dot{m}$.  For the moment, we take $\alpha=0.3$ and
$\beta=0.5$ (i.e. magnetic pressure in equipartition with gas
pressure).  The electron temperature at a given point in the ADAF,
$T_{\rm e}$, can then be determined self-consistently for a given
$\dot{m}$ and $M$ by balancing the heating of the electrons by the
ions against the various radiative cooling mechanisms (within
$r<1000$, it is found that $T_{\rm e}$ is approximately constant in
the range of parameter space of interest and it is $\approx 2-3\times
10^9\K$).  To determine the observed spectrum, we integrate the
emission over the volume and take account of self-absorption effects.
We have taken the inner radius of the disk to correspond with the
innermost stable orbit around a Schwarzschild black hole, $r_{in}=3$,
and the outer radius to be $r_{out}=10^3$ (for which the
characteristic 2-temperature ADAF solution is mantained).

In Fig.~3 we show the spectrum of the advection-dominated disk for
$m=6.3\times 10^8,10^9,1.6\times 10^9$ where $m$ is defined as $m=M/\Msun$
 and $dot{m}$ given by the Bondi accretion rate in eqn. (1).  
The peak in the radio band is due to synchro-cyclotron
emission by the electrons in the magnetic field of the
plasma. The X-ray peak is due to thermal bremsstrahlung.  The
power-law emission extending through the optical band is due to
Comptonization of the synchro-cyclotron emission: more detailed
calculations show this emission to be composed of individual peaks
corresponding to different orders of Compton scattering.  The
positions at which the synchrotron and bremsstrahlung peaks occur and
their relative heights depend on the parameters of the model.  The
synchrotron radiation is self-absorbed and locally gives a Rayleigh Jeans
spectrum, up to a critical frequency, $\nu_c$.  
The bremsstrahlung peak occurs at the thermal frequency $\nu\sim
k_{\rm B}T_e/h$.

 \begin{figure*}
 \centerline{\psfig{figure=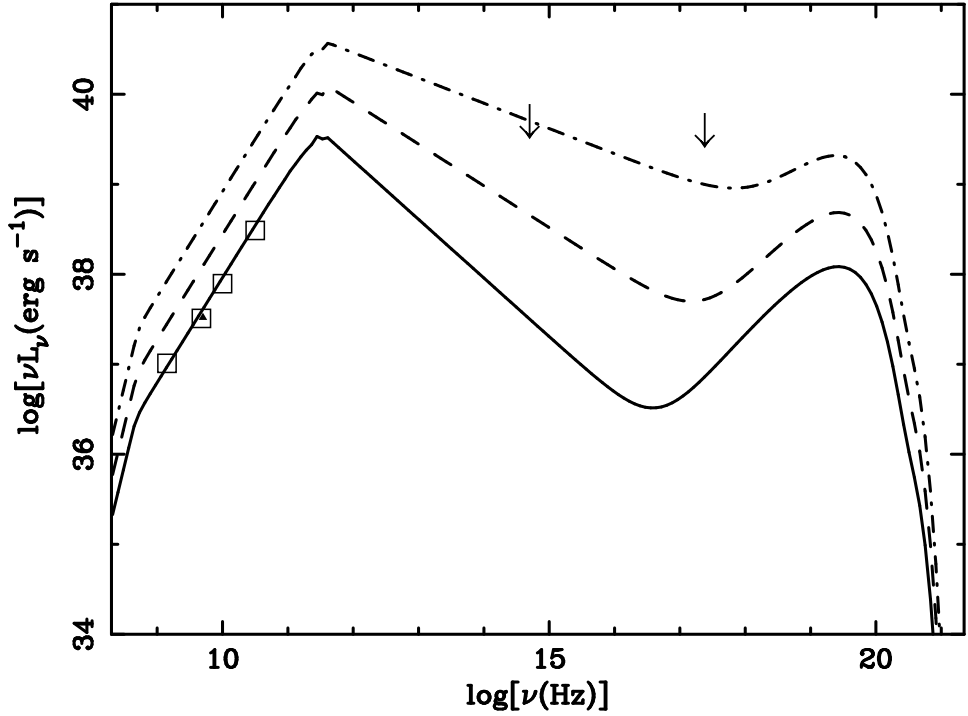,width=0.8\textwidth,angle=270}}
 \caption{Spectrum of M60 calculated with an advection-dominated flow
 extending from $r_{min}=3$ to $r_{max}=1000$. The parameters are
 $\alpha=0.3$, $\beta=0.5$. Three model spectra  are shown:
 for (i) $ m=10^{8.8}-$solid line, (ii) $ m=10^{9.0}$$-$dashed 
 line, (iii) $m=10^{9.2}$$-$dot-dashed line and their corresponding ${\dot m}$ 
given by Bondi accretion formula, eqn.(1).  The 
 squares represent the radio measurements and the arrows the 
upper bounds for the optical and X-ray core emission of M60
 taken from the different references explained
 in the text.  For M60, a distance of $15.8\Mpc$ is assumed.} 
 \end{figure*}

\section{The data versus the ADAF model}
Fig. 3 shows the data on M60 with predicted spectra  
for different black
hole masses and  the relative Bondi accretion rates.
A strong feature of an advection--dominated solution is that the whole spectrum
can be determined by few parameters. With $\alpha$ and $\beta$ given
the only two parameters left are $m$ and $\dot{m}$. In Paper I we had an 
independent determination for the black hole mass, $m$, of M87. The 
only parameter to be determined was
therefore the accretion rate. A comparison with the observational 
limits  confirmed  that the physically 
consistent accretion rate for such a system was indeed the Bondi rate.
Here we therefore use the Bondi formula and vary the mass $m$. We
find that the radio bounds on Fig. 3 allow black hole masses 
$m\approx 10^{9}$, similar to that required by independent arguments for dead
quasars. 
For such masses, the low luminosity of the core of M60 is consistent 
with the data at different
wavelengths only if the accretion involves an ADAF.
Interestingly, the radio data points reproduce the slope ($\propto \nu^{2/5}$)
of the self-absorbed synchrotron radiation (Mahadevan 1996) very well. 
Such agreement
is even more evident than in M87 (Paper I) where contributions from
the powerful jet had to be taken into account. 
It is important to point out that radio observations (Slee et al. 1994)
of many bright elliptical galaxies typically show rising spectra with an
average radio spectral index of $1/3$. 

\begin{figure}
 \centerline{\psfig{figure=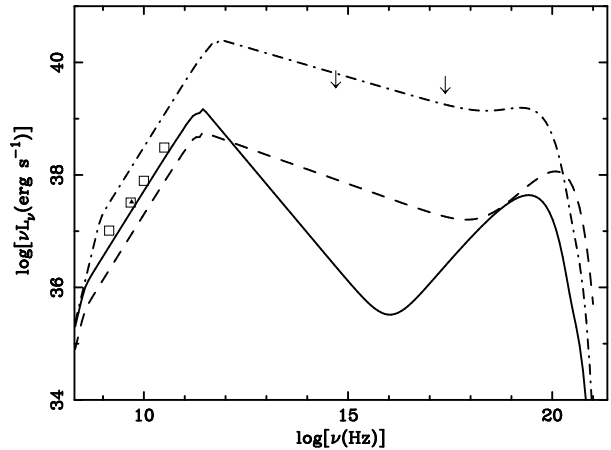,width=0.5\textwidth,angle=270}}
 \caption{Spectra of M60 calculated for  $m=10^{8.8}$. The effects of varying 
$\alpha$ and $\beta$ are shown. 
 (i) the dashed line is obtained for $\alpha=0.3$ and $\beta=0.95$, (ii) the
solid line for  $\alpha=0.5$ and $\beta=0.5$ and (iii) the dot-dashed line 
for  $\alpha=0.1$ and $\beta=0.5$}. 
\end{figure}   

\section{Summary}
 
We have shown that if M60 consists of an advection-dominated system
 accreting from the hot interstellar medium
at a rate determined by the Bondi formula,
then the upper limit on nuclear black hole mass is much higher,
$M\approx10^9\Msun$, than that determined by Fabian \& Canizares (1988), 
$M\le~$few$ \times 10^7\Msun$. 

The poor efficiency of this advective accretion mode accounts 
for the low luminosity observed
 and allows for
 black hole masses of the order of $10^9\Msun$ consistent with the idea 
that bright ellipticals 
like M60, host dead quasars.  

This conclusion is strengthened by the previous 
application of a similar model to the well known active nucleus in M87, 
which has a nuclear black hole mass of $3\times10^9\Msun$ and luminosity
several orders of magnitude fainter than that expected from the Bondi
solution  (Paper I).

An important test for this model would be high frequency radio/submm
  observations. These should reveal an abrupt spectral turnover
above the inner self-absorbed frequency ($\sim 3\times10^{11} {\rm Hz}$)  
predicted by the model. Observations of the turnover would
constrain the allowable viscosity and magnetic parameters $\alpha$ and
$\beta$. Fig. 4 illustrates how different combinations of $\alpha$ and $\beta$
are either ruled out by the available data or  
constrained by the higher frequency observations. In particular, we 
suggest that $\alpha$ and $\beta$ should be anticorrelated for 
a given mass in order to have 
agreement with observations. This means that for relatively
 low values of $\alpha$  ($\sim 0.1$ within the context of ADAF)
  require magnetic fields below their equipartition value 
(i.e. $\beta>0.5$), for a black hole mass $\sim 10^9 \Msun$.

High radio frequency polarization and interferometry  studies 
could indicate and resolve the torus 
perpendicular to the jet.
Better constraints on the X-ray flux will only be obtained with the improved
 resolution of AXAF. 
\section*{Acknowledgements}

TDM acknowledges PPARC and Trinity College, Cambridge, for
support.  ACF thanks the Royal Society for support.  
The  XIMAGE software package has been used for the data analysis.

\end{document}